\documentclass[runningheads,a4paper]{llncs}

\usepackage{amssymb}
\usepackage{graphicx}

\begin{document}

\mainmatter
\title{Riccati-regularized Precision Matrices for Neuroimaging}
\titlerunning{Riccati regularized Precision Matrices for Neuroimaging}

\author{Nicolas Honnorat \and Christos Davatzikos}
\authorrunning{}
\institute{Department of Radiology, University of Pennsylvania, \\ 3700 Hamilton Walk, Richards Building, 7th Floor \\
Philadelphia, PA 19104, USA 
}
\maketitle

\begin{abstract}

The introduction of graph theory in neuroimaging has provided 
invaluable tools for the study of brain connectivity. 
These methods require the definition of a graph, which is 
typically derived by estimating the effective connectivity 
between brain regions through the optimization of an ill-posed inverse problem. 
Considerable efforts have been devoted to the development of methods 
extracting sparse connectivity graphs. 

The present paper aims at highlighting the benefits of an alternative approach. 
We investigate low-rank L2 regularized matrices recently introduced 
under the denomination of Riccati regularized precision matrices. 
We demonstrate their benefits for the analysis of cortical thickness map and for the extraction of functional biomarkers from resting state fMRI scans. In addition, we explain how speed and result quality can be further improved with random projections. 
The promising results obtained using the Human Connectome Project dataset as well as the numerous possible extensions 
and applications suggest that Riccati precision matrices might usefully complement current sparse approaches.
\keywords{rs-fMRI,precision,sparse inverse covariance}
\end{abstract}

\section{Introduction}


Resting-state functional MRI (rs-fMRI) studies of brain connectivity have received a considerable amount of interest. 
Thanks to the continuous improvement of imaging technics and the development of big data infrastructures, 
large datasets are now available for conducting these researches \cite{hcp}. 
An increasing effort has been devoted to the development of mathematical frameworks able to 
distillate these humongous datasets into robust and concise causal models and biomarkers, with the hope of 
better describing cognition, teasing out the mechanisms underlying brain diseases and defining 
novel clinical dimensions \cite{insel}. 
Different measures of connectivity were proposed \cite{varoquaux_review} and 
graph theoretical approaches spread widely \cite{rsfMRIreviewNature}. 
The most straightforward approaches assume that the time series observed during a rs-fMRI scan are 
generated by a multivariate Gaussian process and attempt to analyze its structure. 
Once rs-fMRI have been registered, motion corrected, denoised and normalized, these studies typically 
starts with the definition of locations in the gray matter. The connectivity between 
these nodes is measured by computing the covariance of their time series and a sparse graph is built 
from the inverse of the covariance matrix, also known as precision matrix. 
These last two steps can be performed jointly, by directly estimating sparse precision matrices 
\cite{review_precisions,varoquaux_review,varoquaux_nips}. 
However and despite very impressive recent development, the estimation of sparse precision matrices is still 
time-consuming for large matrices \cite{quic2,review_precisions}.

In this paper, we propose an alternative approach based on low-rank Riccati regularized precision matrices, 
introduced first by Witten and Tibshirani \cite{scout} and formalized by Honorio and Jaakkola \cite{riccati_partial_correlations}. 
As for sparse matrices, we measure network characteristics directly from these matrices \cite{principle_integration_segregation,varoquaux_nips}. 
Our approach offers several benefits such as a very competitive computational efficiency which deteriorates only linearly with data dimension,
straightforward practical and theoretical extensions. 
We demonstrate that reducing the dimension of the input neuroimaging signals via random projections \cite{structure_from_randomness} can simultaneously improve test-retest performances and reduce computational burden and we present two extensions: the estimation of precision matrices at a population level, and the adaptation of Riccati penalties to regions of interests. 
These results were established using the data available for the hundred unrelated subjects of the HCP dataset \cite{hcp}. An in-depth test-retest validation was carried out by reducing the spatial dimension of the resting state fMRI (rs-fMRI) scans with Glasser et al. parcellation \cite{180parcellation}. 
In addition, we demonstrate that our approach is able to handle full resolution data and other modalities 
by analyzing cortical thickness maps.

The remainder of the paper is organized as follows. 
The methods combined in this work are presented in section 2. 
Section 3 presents several variants of our apporach addressing related neuroscience applications. 
The experimental results are presented in section 4 and discussion concludes the paper.

\section{Methods}

The random projection method described in section 2.1 was used as a preprocessing step for reducing the dimensionality of our imaging data 
and filtering noise. We present in section 2.2 a generalization of Honorio and Jaakkola Riccati regularized precision matrices 
\cite{riccati_partial_correlations}. The Gaussian entropy introduced by Tononi, Sporns and Edelman \cite{principle_integration_segregation} 
for measuring functional network integration is presented in section 2.3. We used this measure for extracting biomarkers from 
Riccati regularized precision matrices.

\subsection{Random Projection}

Random projections were proposed for compressing high-dimensional measurement while preserving their Euclidean distance. 
The random projections proposed by Halko, Martinsson and Tropp in \cite{structure_from_randomness} achieve performances close to a truncated 
singular value decomposition (TSVD): when a data matrix $X$ of size $N \times T$, $T < N$ is projected for creating a thinner matrix $Y$ of size $N \times t$, $t < T$, the $t$ non-zero singular values of $Y$ are close to the $t$ largest singular values of $X$. 
These random projections were proposed for accelerating the computation of 
singular value decompositions (SVD) \cite{structure_from_randomness}. 

\cite{structure_from_randomness} is straightforward to implement. 
Figure \ref{fig:rp} provides the pseudo code of the random projection algorithm used for this work. 
This algorithm generates an orthogonal projection matrix by randomly combining matrix rows and orthonormalizing the basis obtained through the Gram-Schmidt process. As explained in \cite{structure_from_randomness} and reported in section \ref{sec:exp:truncation} for the HCP data, the quality of the projection basis can be significantly improved by running a few power iterations, but the computation cost grows rapidly with the number of matrix rows. 
\begin{figure}[h]
\vspace{-1.0em}
\framebox{\parbox[l]{\textwidth}{
\textbf{input}: data $X$ of size $N \times T$, parameter $t$ and number of power iteration $q$ \\
1.~~~form the $N \times t$ matrix $\Omega$ by sampling from Gaussian distribution ${\cal N}(0,1)$ \\
2.~~~$U \leftarrow \left(X^T X\right)^q X^T \Omega $ \\
3.~~~get $W$ by orthonormalizing the columns of $U$ with Gram-Schmidt process \\
4.~~~$Y\leftarrow X W$ \\
\textbf{output}: projected data $Y$, of size $N \times t$
}}
\caption{\label{fig:rp}Random projection for dimensionality reduction \cite{structure_from_randomness}.}
\vspace{-3.0em}
\end{figure}

\subsection{Riccati regularized Precision Matrices}
\label{sec:method:riccati}
Let $X$ denote a matrix of size $N \times T$ containing $N$ time series of $T$ time points, normalized to zero mean and unit variance. 
Let denote the associated covariance matrix with $C=\frac{1}{T}XX^{T}$. 
Sparse, Tikhonov and Riccati regularized precision matrices are obtained by solving the following optimization problem:
\begin{equation}
\label{formula:optimization}
\texttt{argmax}_{Q \succ 0} \left[ \texttt{log det}Q - \langle C,Q \rangle - \rho R(Q) \right]
\end{equation}
where an L1 norm is chosen for $R$ for generating sparse precision matrices \cite{varoquaux_nips}, the trace of $Q$ in the case of Tikhonov regularization \cite{riccati_partial_correlations}, and $R$ is the square of the Frobenius norm for Riccati regularized precision matrices \cite{riccati_partial_correlations}. As explained in \cite{scout,riccati_partial_correlations}, Riccati regularization is a ridge penalty on the components of the precision matrix whereas, when precision matrices are computed for solving a linear regression, Tikhonov regularization corresponds to a ridge penalty on the coefficients of a linear regression. 
In this work, we generalize the Riccati regularization described in \cite{scout} and \cite{riccati_partial_correlations} 
by introducing an invertible matrix $V$ and working with the penalty:
\begin{equation}
R(Q)=\frac{1}{2}|| VQV^T ||^2_2
\end{equation}
When V is a diagonal matrix, $V=diag(v)$, the penalty $R$ becomes a squared weighted Frobenius norm, that can be expressed as follows:
\begin{eqnarray}
R(Q)&=&\frac{1}{2}|| B \odot Q ||^2_2 \\
B &=& v v^T
\end{eqnarray}
where $\odot$ denotes the Hadamard product. This specific case, easy to interpret and interesting for applications, will be referred as Hadamard-Riccati regularization.

An analytical solution of (\ref{formula:optimization}) is obtained by following Honorio and Jaakkola derivation \cite{riccati_partial_correlations}, which bears similarity with the derivation of the $Scout(2,.)$ method of Witten and Tibshirani \cite{scout}. 
More precisely, the extrema of the objective (\ref{formula:optimization}) are found by solving:
\begin{eqnarray}
Q^{-1} - C - \rho V^T V Q V^T V &=& 0
\end{eqnarray}
Following \cite{riccati_partial_correlations}, $Q$ is obtained as a matrix geometric mean: 
\begin{eqnarray}
V Q V^T &=& P = \left(\frac{1}{\rho}D\right) \# \left(D^{-1}+\frac{1}{4\rho}D\right)-\frac{1}{2\rho}D \\
\texttt{where~}D &=& V^{-T} C V^{-1} = (\frac{1}{\sqrt{T}}V^{-T}X)(\frac{1}{\sqrt{T}}V^{-T}X)^T
\end{eqnarray}
According to the properties of geometric means of matrices $\#$, recalled in \cite{lim2006}, the eigenvectors of $D$ are also eigenvectors of $P$, and an eigenvalue $p$ of $P$ depends only on the eigenvalue $d$ of $D$ associated with the same eigenvector:
\begin{equation}
\label{eqn:finale}
p(d)=\sqrt{\frac{d}{\rho}\left(\frac{1}{d}+\frac{d}{4\rho}\right)}-\frac{d}{2\rho}
\end{equation}

This property leads to the efficient computation of $Q$ presented in figure \ref{alg:riccati}. 
The computation of Hadamard-Riccati regularized precision matrices, for which matrices $V$ are diagonal, is almost as fast as the original \cite{riccati_partial_correlations}. 
We found that random projections \cite{structure_from_randomness} are of prominent interest for the computation of Riccati regularized inverses. 
First, because they accelerate all the computations by reducing matrices dimensions. 
Second, because they provide a direct control of the rank of the rank-deficient part of the precision matrix $Q$.
And third because they reduce precision matrices noise by truncating the small singular values of the covariance matrices $C$. 
This protective effect is illustrated in section \ref{sec:exp:truncation}.

\begin{figure}
\vspace{-1.0em}
\framebox{\parbox[l]{\textwidth}{
\textbf{input}: time series $X$ of size $N \times T$, parameter $\rho$ and $N \times N$ invertible matrix $V$ \\
1.~~~$[U,S,Z] \leftarrow {SVD} \left( \frac{1}{\sqrt{T}}V^{-T}X \right)$; $U$ left singular vectors, $S$ contains singular values \\
2.~~~form diagonal matrix $\Omega=diag(\omega_1, .. , \omega_n)$ from singular values $s_1, .., s_m$: \\
$~~~~~~~~~\omega_i = \sqrt{\frac{1}{\rho}+\frac{s_i^4}{4\rho^2}}-\frac{s_i^2}{2\rho}-\frac{1}{\sqrt{\rho}}$ \\\
3.~~~$W\leftarrow V^{-1}U$ \\
4.~~~$Q\leftarrow W\Omega W^T +\frac{1}{\sqrt{\rho}}V^{-1}V^{-T}$ \\
\textbf{output}: Riccati-penalized precision matrix $Q$
}}
\caption{\label{alg:riccati}Computation of Riccati-penalized precision matrix $Q$, for input time series $X$, the penalization $\rho$ and a $N \times N$ invertible matrix $V$.}
\vspace{-3.0em}
\end{figure}

\subsection{Tononi-Sporns-Edelman Entropy}

\begin{figure}
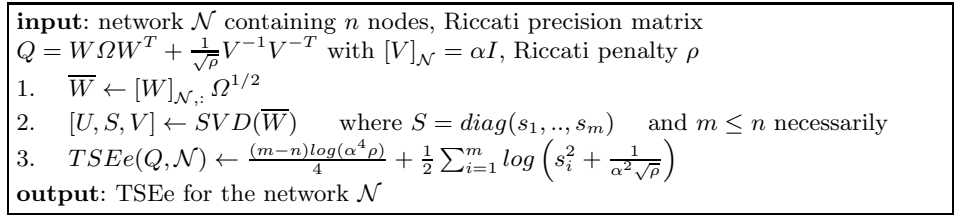

\vspace{-1.0em}
\framebox{\parbox[l]{\textwidth}{
\textbf{input}: network ${\cal N}$ containing $n$ nodes, Riccati precision matrix \\ $Q=W\Omega W^T +\frac{1}{\sqrt{\rho}}V^{-1}V^{-T}$ with $\left[V\right]_{\cal N}=\alpha I$, Riccati penalty $\rho$ \\
1.~~~ $\overline{W} \leftarrow \left[W\right]_{\cal N,:}\Omega^{1/2}$ \\
2.~~~ $\left[U,S,V\right] \leftarrow SVD(\overline{W})$ ~~~~where $S=diag(s_1,..,s_m)$ ~~ and $m\leq n$ necessarily \\
3.~~~ $TSEe(Q,{\cal N}) \leftarrow \frac{(m-n)log(\alpha^4\rho)}{4}+\frac{1}{2}\sum_{i=1}^{m}{ log \left(s_i^2+\frac{1}{\alpha^2 \sqrt{\rho}}\right) } $\\
\textbf{output}: TSEe for the network ${\cal N}$ 
}}
\caption{\label{alg:fastTSEe} Efficient computation of the TSEe for constant Riccati penalties.}
\vspace{-2.0em}
\end{figure}

Tononi, Sporns and Edelman introduced a measure of functional integration derived from precision matrices \cite{principle_integration_segregation}. 
This measure will be referred as Tononi-Sporns-Edelman entropy (TSEe) in the sequel. 
Under the standard assumption that functional time series are Gaussian, TSEe measures the Gaussian entropy of a functional networks ${\cal N}$ as follows: 
\begin{equation}
TSEe(Q,{\cal N})=\frac{1}{2}log det \left(\left[Q\right]_{\cal N}\right)
\end{equation}
where $\left[Q\right]_{\cal N}$ denotes the restriction of the precision matrix $Q$ to the nodes in the network ${\cal N}$. 
TSEe is a standard measure of functional integration and has already been used for measuring the integration of the networks 
derived from sparse precision matrices \cite{varoquaux_nips}. 
In this work, TSEe was measured for Riccati regularized precision matrices as well. 
When the penalty $R$ is constant over the network ${\cal N}$, the structure of the Riccati precision matrix can be exploited for accelerating TSEe computation. A constant penalty $R$ over ${\cal N}$ corresponds indeed to a matrix $V$ proportional to the identity for the nodes in ${\cal N}$:
\begin{equation}
\left[V\right]_{\cal N}=\alpha I
\end{equation}
Under this assumption and following the notations of figure \ref{alg:riccati}: 
\begin{equation}
\left[Q\right]_{\cal N}=\left[W\right]_{\cal N,:}\Omega \left[W\right]^{T}_{\cal N,:}+\frac{1}{\alpha^2 \sqrt{\rho}} I
\end{equation}
where $\left[W\right]_{\cal N,:}$ denotes the restriction of the rows of $W$ to the nodes in the network ${\cal N}$. Because $\Omega$ is a diagonal matrix with strictly positive diagonal components the following matrix can be computed in a single pass over $\left[W\right]_{\cal N,:}$:
\begin{equation}
\overline{W}=\left[W\right]_{\cal N,:}\Omega^{1/2}
\end{equation}
Let $s_i$ denote one of the $m$ singular values of $\overline{W}$ and $n$ the number of nodes of the network ${\cal N}$. The left singular vector of $\overline{W}$ associated to $s_i$ is an eigenvector of $\left[Q\right]_{\cal N}$ and the associated eigenvalue $\lambda_i$ is equal to $s_i^2+\frac{1}{\alpha^2 \sqrt{\rho}}$. 
The remaining eigenvectors of $\left[Q\right]_{\cal N}$ are associated with the same eigenvalue: $\frac{1}{\alpha^2 \sqrt{\rho}}$.
As a result, TSEe can be computed as explained in figure \ref{alg:fastTSEe}, at the cost of a single SVD.

\section{Applications}


\subsection{Robust Structural Distances}
\label{method:distance}

Cortical thickness (CT) is a scalar measured from structural MRI describing local cortical gray matter geometry \cite{structural_covariance_review,freesurfer}. 
The structural covariance matrix is obtained by computing CT covariance across a population, for all pairs of brain locations. 
Large structural covariances indicate that brain regions develop, age or suffer from a disease in a similar way across a population \cite{structural_covariance_review}. 
The inverse of a structural covariance $C_{\cal S}$ obtained for a healthy population can be used for defining a Mahalanobis distance $d_{\cal S}$ teasing out abnormal CT maps: 
\begin{equation}
d_{\cal S}(a,b)=\sqrt{(a-b)^{T}C^{-1}_{\cal S}(a-b)}
\end{equation}
This distance is indeed small when the difference between CT maps $a$ and $b$ is likely to be observed in the healthy population, whereas 
large distances correspond to unusual CT variations. 
In this work, we introduce Riccati regularized structural precision matrices. We show experimentally in section \ref{exp:distance} that regularization and random projections improve structural distance robustness.

\subsection{Shared Functional Networks}
\label{method:shared}

An increasing effort has been dedicated to the extraction of biomarkers capturing the specificities of 
individual rs-fMRI scans, with the aim of developing rs-fMRI based diagnostic tools. 
Because these scans are strongly affected by noise and subject motion, 
several regularization strategies were proposed such as the introduction of population averages \cite{varoquaux_nips}. 

In this work, instead of introducing a group average precision matrix and penalizing the differences between individual scans and the group average \cite{varoquaux_nips} we propose to perform a joint SVD (JSVD) when computing Riccati regularized precision matrices. 
This JSVD forces Riccati regularized precision matrices to share their eigenvectors. As a result, scan specificities are encoded in a reduced set of values, the eigenvalues of the Riccati regularized precision matrices, which can be interpreted as scan-specific loadings. 
This modeling offers many advantages for investigating neurodevelopment, aging and brain diseases \cite{varoquaux_nips,harini_IPMI}. 
The shared eigenvectors will be referred as shared functional networks in the sequel.



\subsection{Functional Network Biomarkers}
\label{method:biomarkers}

TSEe is an interesting functional biomarker. However, when small brain networks are investigated 
TSEe might be corrupted by a noise induced by the random variation of the other components of the precision matrix. 
We propose to address this issue by penalizing more the components of the precision matrix corresponding to nodes outside the network of interest. 
We design a simple penalty achieving this goal by choosing for $V$ a diagonal matrix equal to the identity when restricted to the nodes of the network and to $\alpha$ times the identity, $\alpha>1$, when restricted to the other nodes. 
As explained in section \ref{sec:method:riccati} such a penalty is an Hadamard-Riccati penalty. 
When $\alpha$ is increased, this penalty gradually isolates the network of interest from the rest of the brain. As illustrated in section \ref{exp:biomarkers}, this effect can improve test-retest performances for some functional networks.


\section{Experimental Validation}

\subsection{HCP Dataset}
\label{sec:exp:truncation}

All the experiments presented in this work were carried out with the hundred unrelated subjects 
of the HCP dataset \cite{hcp}. 
For each subject, four 15 minutes long rs-fMRI scans of 1200 time points are available, and several maps 
describing the local geometry of the cortex such as cortical thickness \cite{hcp_preprocessing}.
We used the rs-fMRI scans processed with the ICA+FIX pipeline with MSMAll registration and the cortical thickness map 
registered in the 32k Conte69 atlas, also registered with MSMAll \cite{hcp_preprocessing}.
Rs-fMRI scans were bandpass-filtered between $0.05$ and $0.1$ Hz by an equiripple finite impulse response filter and the first two hundred timepoints impacted by the temporal filtering were discarded. Cortical thickness maps outliers were discarded by thresholding each map independently at $\pm 4.4478$ median absolute deviation from the median. This thresholding can be interpreted as a counterpart of the standard thresholding of Gaussian variable to three standard deviations from the mean robust to the presence of outliers \cite{mad}. All the time series and concatenated cortical thickness maps were normalized to zero mean and unit variance. 
The spatial dimension of the data was reduced by averaging the neuroimaging signals over Glasser et al. multi-modal parcellation \cite{180parcellation}. 
The hundred eighty time series obtained for each hemisphere were normalized again to zero mean and unit variance.

\begin{figure}[h]
\vspace{-1.0em}
(a)\includegraphics[width=0.28\linewidth]{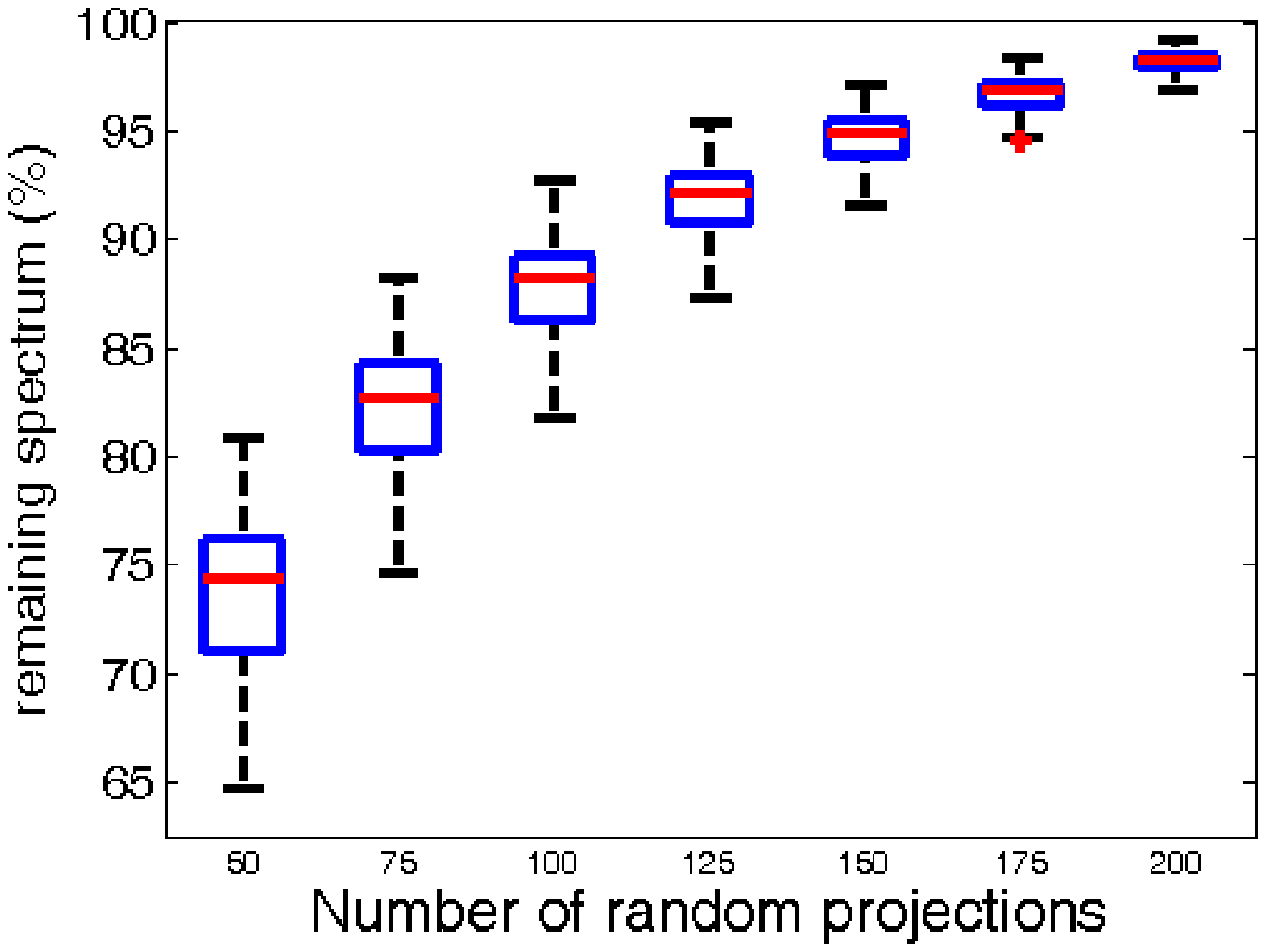}
(b)\includegraphics[width=0.28\linewidth]{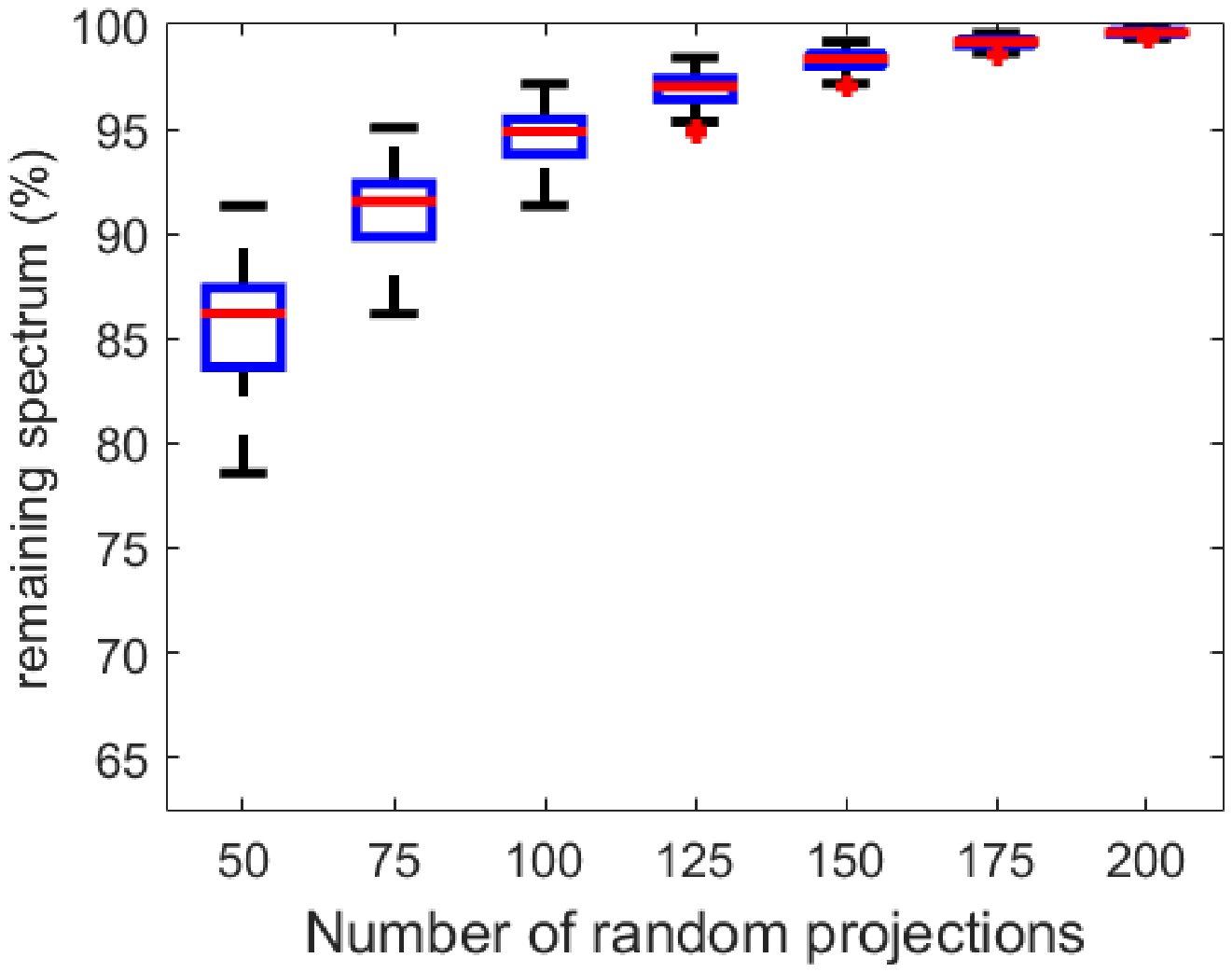}
(c)\includegraphics[width=0.31\linewidth]{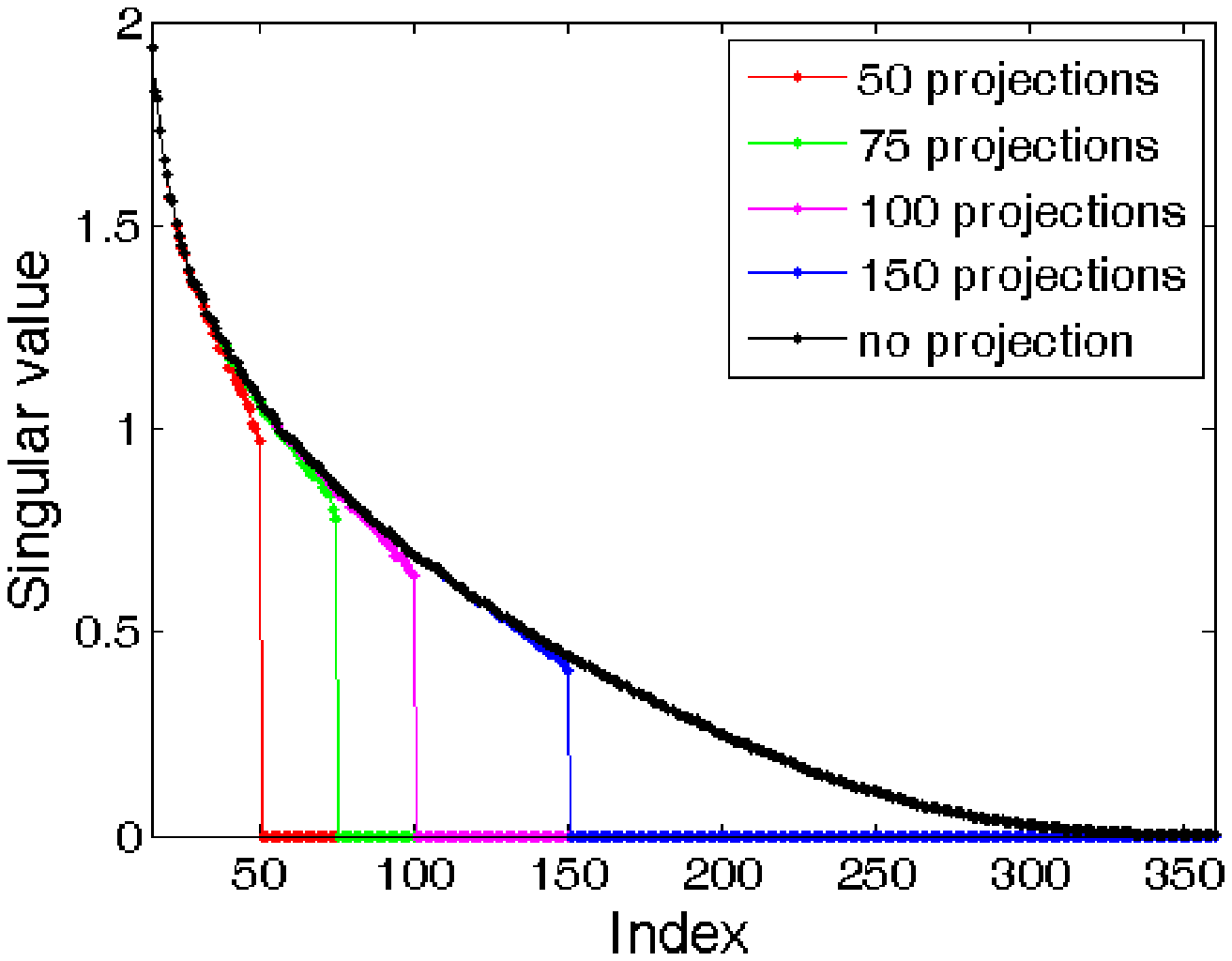}
\caption{\label{fig:exp:rp} \textbf{RP for the functional data}. Remaining spectrum for increasing number of random projections 
(a) without power iterations (b) three power iterations ($q=3$) (c) For the first subject: singular values before and after random projections ($q=3$).}
\end{figure}

The quality of the random projections (RP) was estimated by (1) concatenating the four rs-fMRI scans of each subject, (2) measuring for each subject the proportion of the squared Frobenius norm of the signal kept by the random projections, and (3) comparing the singular values of the time series before and after random projections. The results presented in figure \ref{fig:exp:rp} demonstrate that RP behave almost like a perfect truncated SVD (TSVD) after only three power iterations. 
The results suggest also that the 4000 timepoints time series can be randomly projected into a dimension 200 with negligible information loss, even without power iterations. 

\subsection{Robust Structural Distances}
\label{exp:distance}

\begin{figure}[h]
\begin{minipage}{0.66\linewidth}
(a)\includegraphics[width=0.95\linewidth]{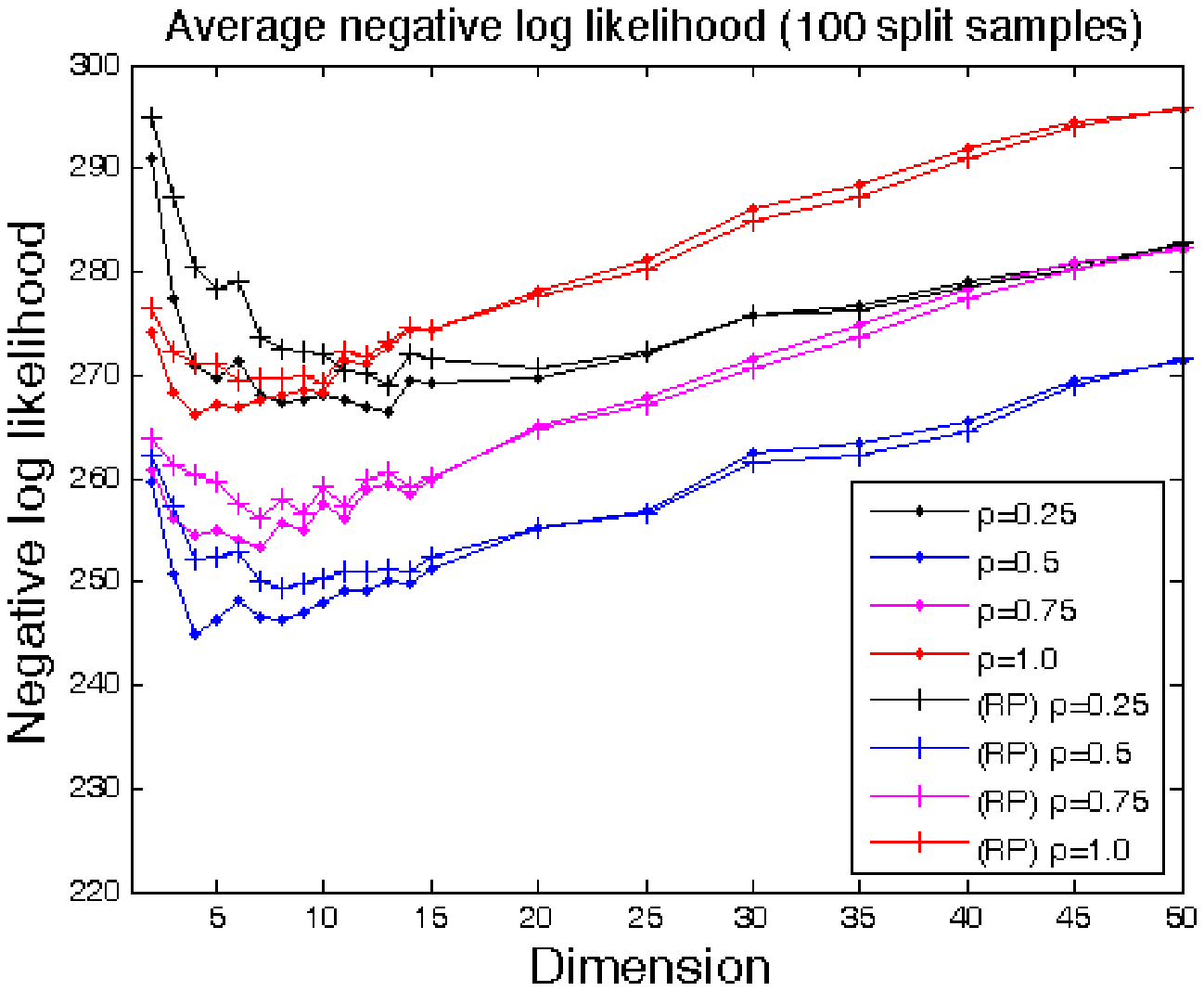}
\end{minipage}
\begin{minipage}{0.32\linewidth}
\includegraphics[width=1.0\linewidth]{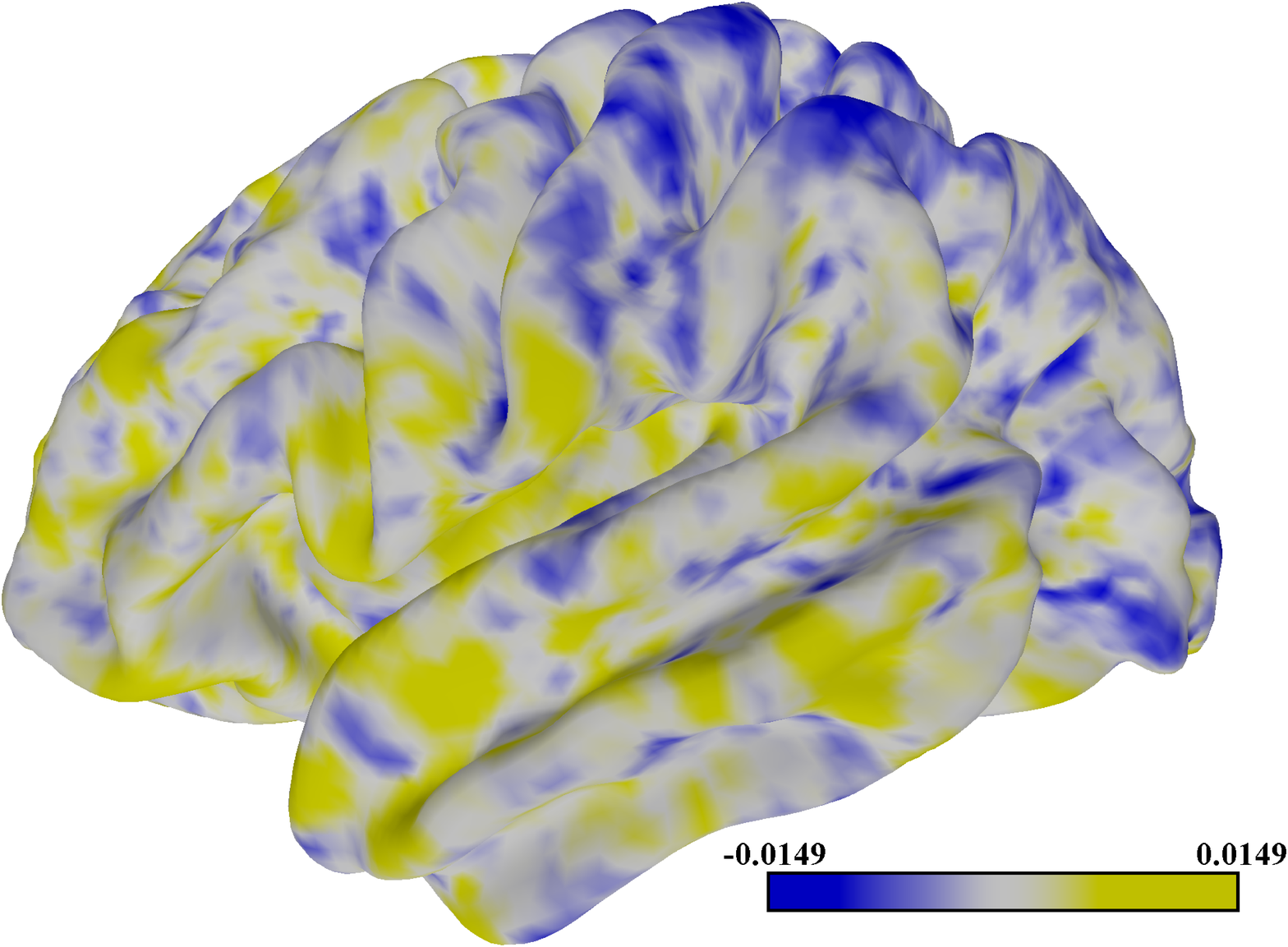}
(b)\includegraphics[width=0.9\linewidth]{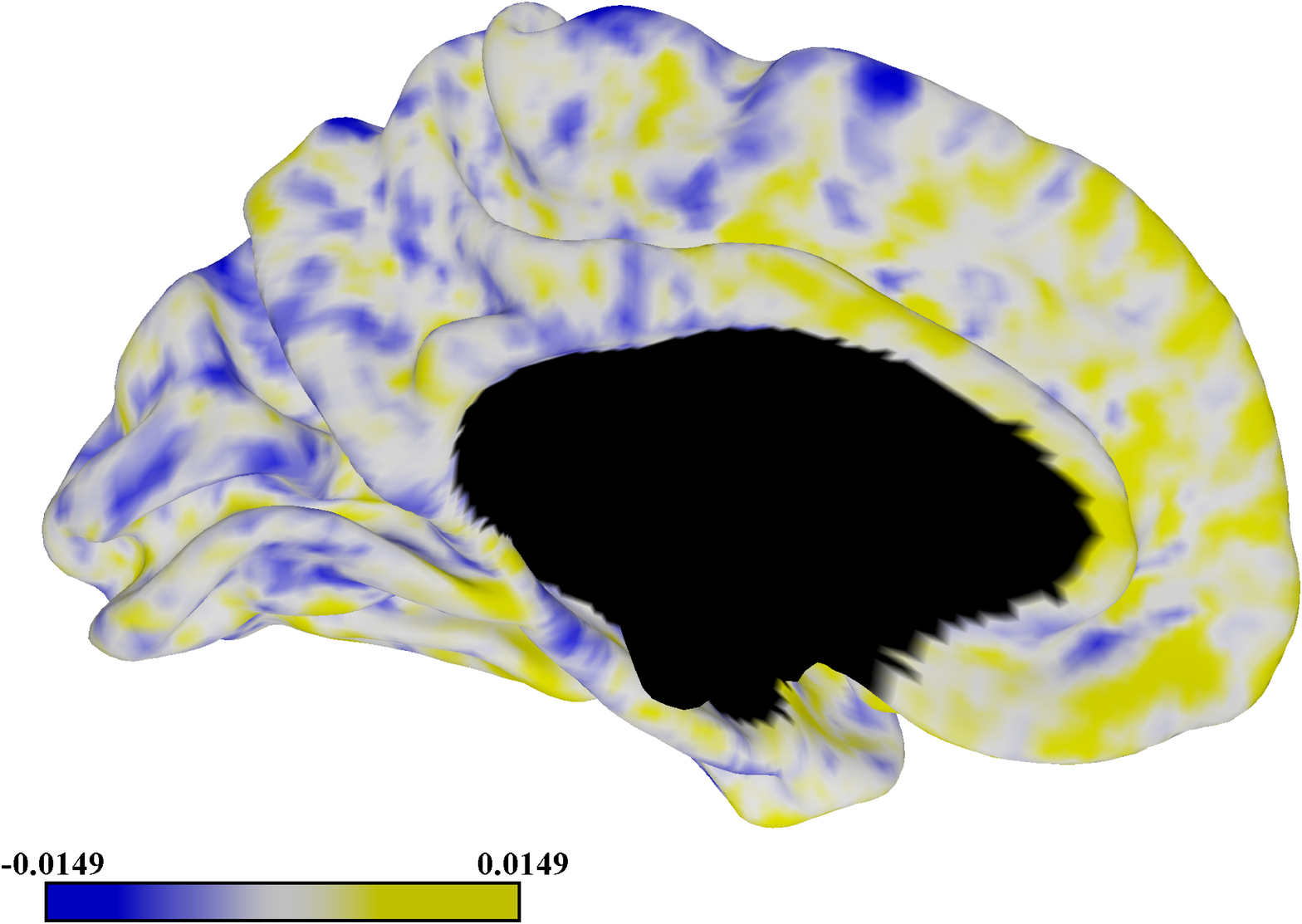}
\end{minipage}
\caption{\label{exp:distance} \textbf{Structural precision}
(a) Average split sample negative log-likelihood (100 repetitions) of Riccati regularized precision matrices built for the cortical thickness (CT) averaged over the 180 parcellation, with respect to the ``dimension'': the number of singular values kept by the truncated SVD or by the random projection (RP). Dimension 50 corresponds to the original data, without RP or TSVD. Four different Riccati penalties $\rho$ were tested. (b) one of the seven modes of CT variation obtained at full brain resolution for the left hemisphere, $\rho=100.0$ and RP into seven dimensions. This map corresponds to a column of the matrix $W$ defined in Fig.(\ref{alg:riccati}).
}
\vspace{-1.0em}
\end{figure}

Riccati regularized structural precision matrices reliability was measured by the split sample negative log likelihood, a measure decreasing with reproducibility. More precisely, the dataset was randomly split a hundred time into two groups of fifty subjects. For each split, the CT maps of the two groups were concatenated and normalized to zero mean and unit variance separately. A precision matrix $Q$ was computed for the first group and its negative log likelihood was measured by:
\begin{eqnarray}
NLL(Q)= \langle C,Q \rangle - \texttt{log det}Q
\end{eqnarray}
where $C$ is the structural covariance obtained from the second group. This test-retest procedure estimates the ability of the precision matrices learned for the first group to fit/generalize to the remaining HCP subjects. 
The results reported in figure \ref{exp:distance}.a demonstrate that the reliability of structural precision matrices is improved by TSVD and RP and reaches an optimum at small dimension and for a moderate penalty $\rho=0.5$. RP and TSVD results are very close, for large dimensions and large penalties. For the sake of simplicity, $V$ was set to the identity for these experiments.

We measured the ability of our method to handle large data by computing structural precision matrices at full Conte69 32k atlas resolution and both hemisphere simultaneously (59412 nodes total). On a standard office computer running an Intel Core i5-200 CPU 3.3 GHz and 8Gb RAM, without random projections, the Riccati precisions were obtained in $12.47$ seconds on average (over 100 runs). 
A random projection to dimension seven followed by the computation of the Riccati precision required $0.28$ seconds on average (over 100 runs) and captured CT variation modes similar to the one presented in figure \ref{exp:distance}.b. 
By comparison, sparse precision matrices are typically obtained in two hours for 20000 nodes without GPU acceleration \cite{quic2}.

\subsection{Shared Functional Networks}
\label{exp:shared}

\begin{figure}[h]
(a)\includegraphics[width=0.40\linewidth]{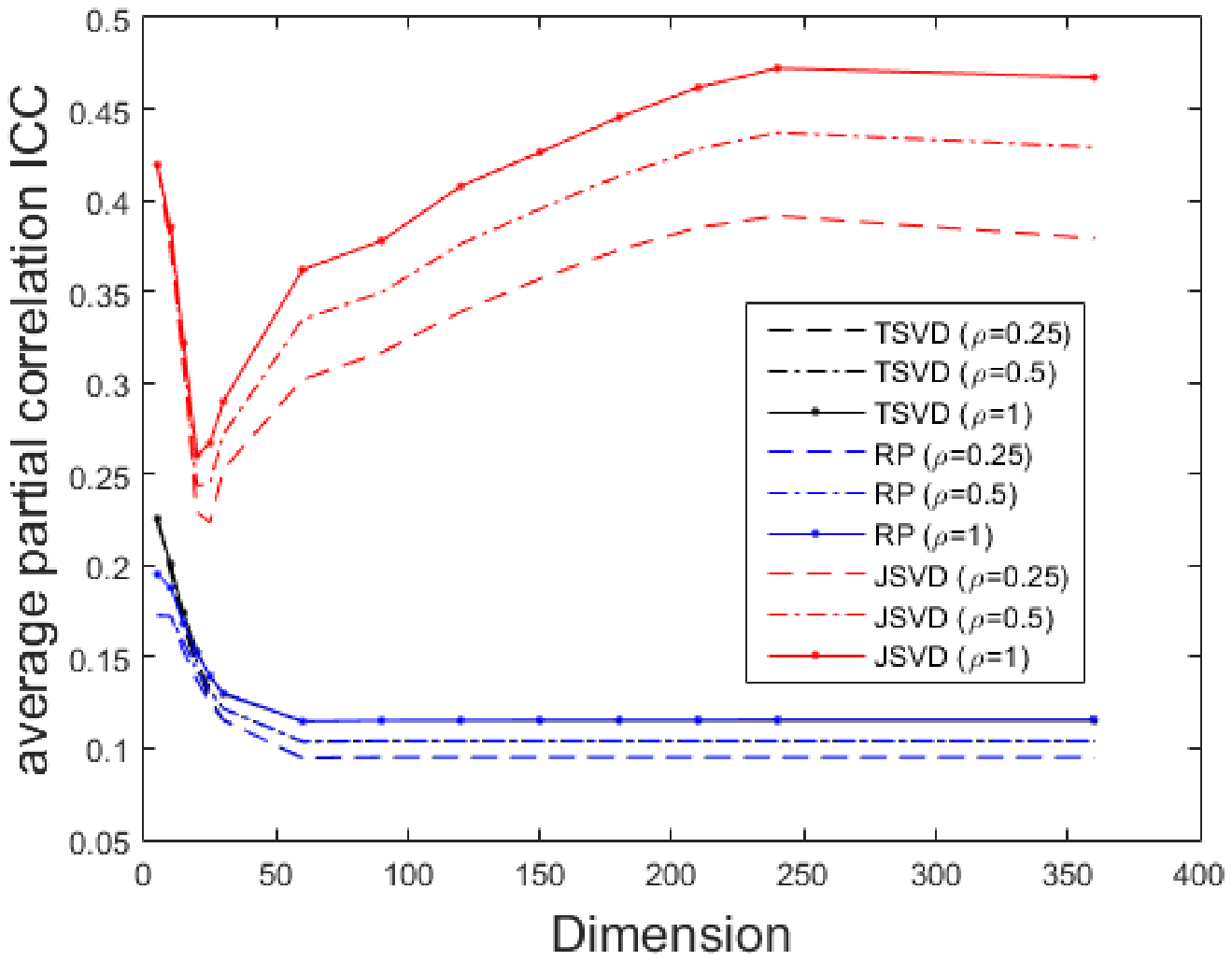}
(b)\includegraphics[width=0.40\linewidth]{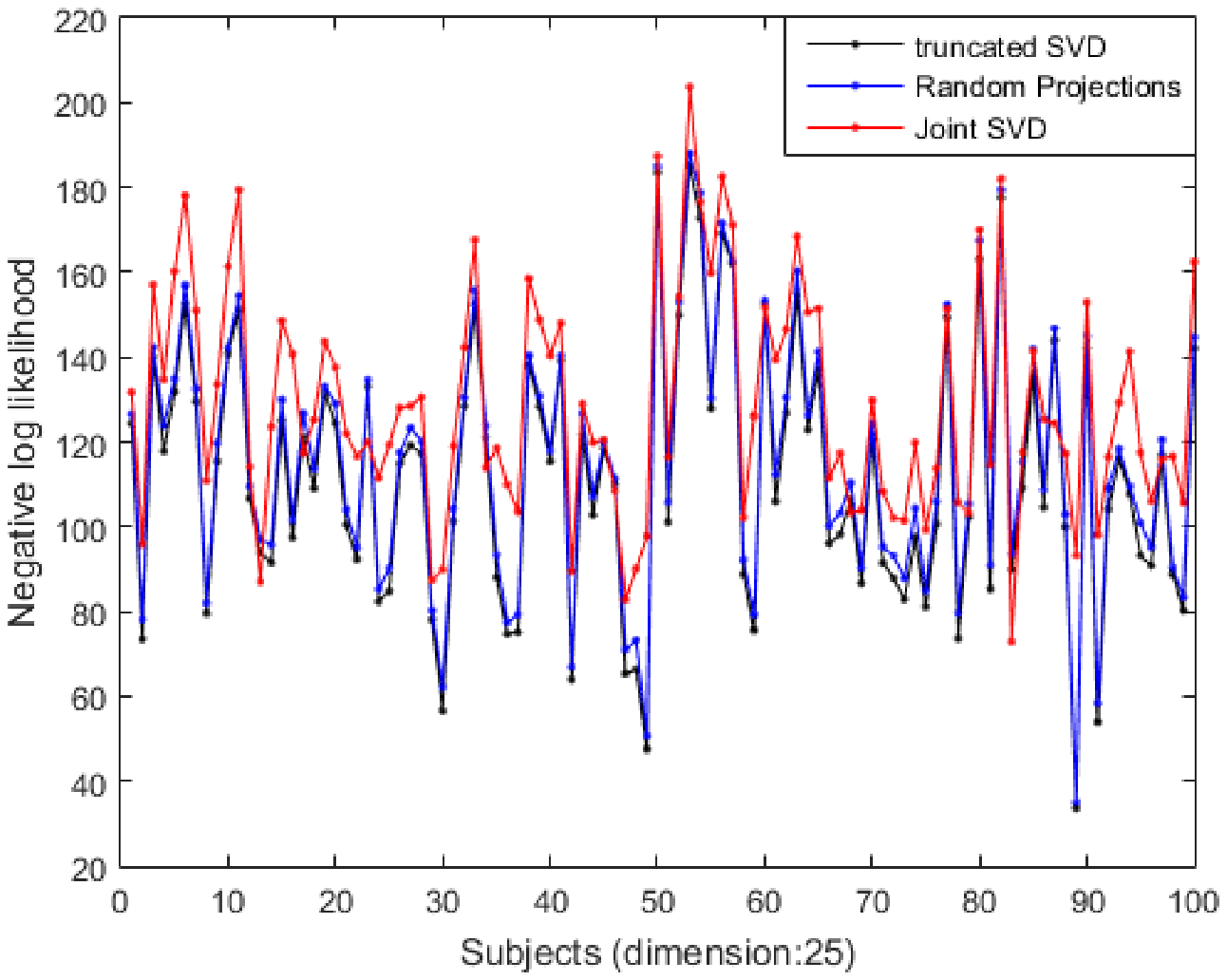}
\caption{\label{fig:exp:shared} \textbf{Shared functional networks better capture subject specificities but generalize slightly less.} 
(a) average ICC observed for the partial correlations derived from the Riccati precision matrices of the entire dataset, for different dimensions and penalties $\rho$. TSVD and RP results differ only at small dimension.
(b) for each subject: negative log likelihood of the precision matrices obtained for the first two scans of the subject, evaluated with the last two scans. RP and TSVD results are close. JointSVD precisions, obtained for all the subjects simultaneously, generalize slightly less. The dimension was set to 25 and $\rho$ to $0.25$.
}
\vspace{-1.0em}
\end{figure}

The joint SVD (JSVD) method \cite{jointSVD} was used in this work for defining shared functional networks. 
We compared the ability of JSVD, TSVD, and RP to robustly capture individual function by computing first Riccati regularized 
precision matrices for all the rs-fMRI scans of the hundred unrelated HCP subjects, for different dimensions and penalties $\rho$. 
Because functional networks are usually described for correlations or partial correlations, we derived partial correlations from all these precision matrices. 
We compared the methods by measuring the average intraclass correlation coefficient (ICC) of the partial correlations. We mesured thus if the repeated scans of a single subject were producing partial correlation more similar to each other than scans of different subjects. 
We measured an $ICC(C,1)$ \cite{icc}. For the sake of simplicity, $V$ was set to the identity for these experiments.
The results of figure \ref{fig:exp:shared}.a clearly demonstrate that JSVD better captures the specificities of subjects brain function. 

We checked the reliability/reproducibility of JSVD results by concatenating the first two and last two scans of each subject, computing JSVD, TSVD and RP Riccati regularized matrices for the first scans and measuring the negative log likelihood obtained with the last scans. As indicated in figure \ref{fig:exp:shared}.b, we observed that JSVD matrices generalize slightly less than their TSVD and RP counterparts. These results suggest that a given population is much better described using JSVD, at the cost of a small decrease of generalizability to other populations. 

\subsection{Functional Network Biomarkers}
\label{exp:biomarkers}

\begin{figure}[h]
\begin{minipage}{0.32\linewidth}
(a)\includegraphics[width=0.81\linewidth]{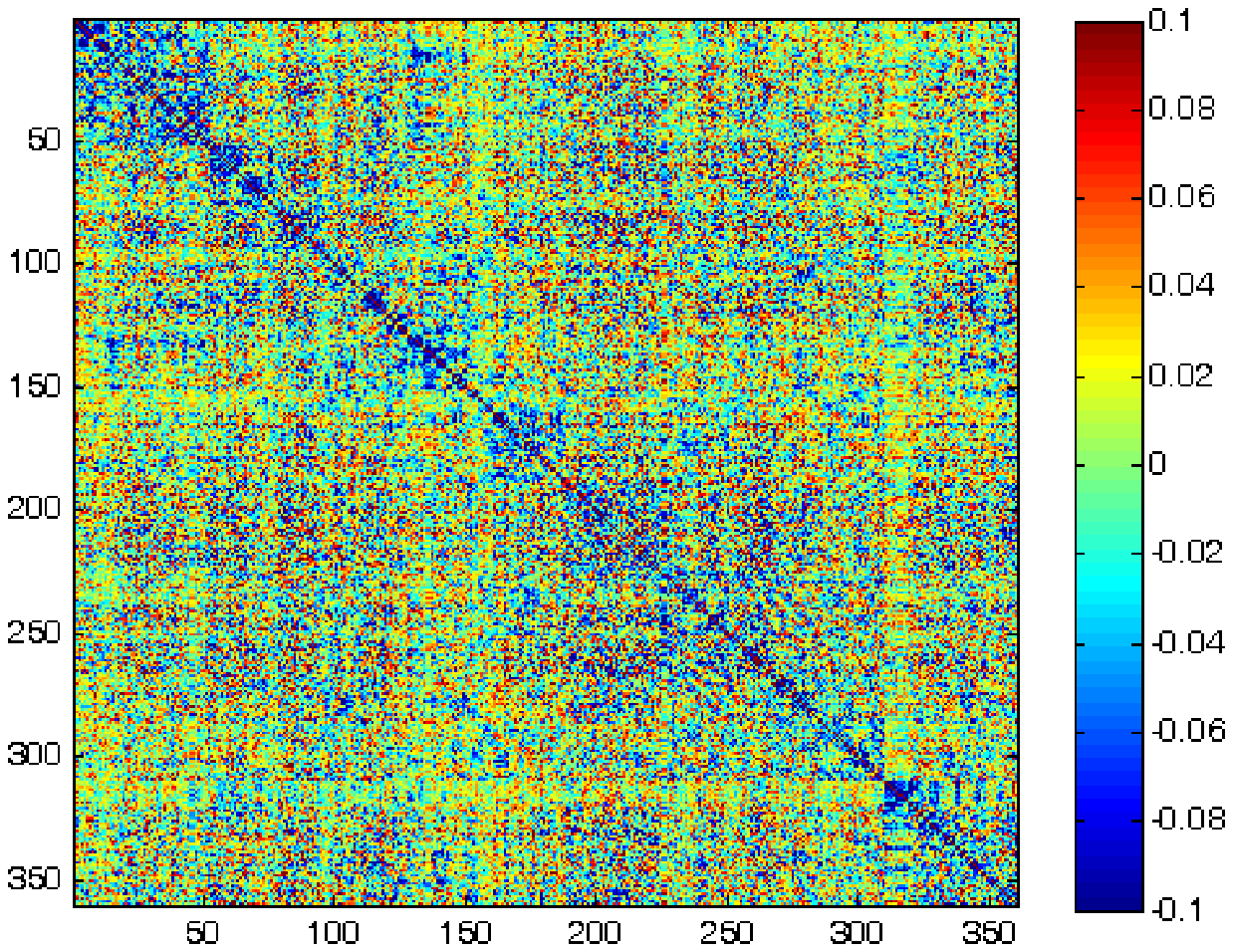}
(b)\includegraphics[width=0.81\linewidth]{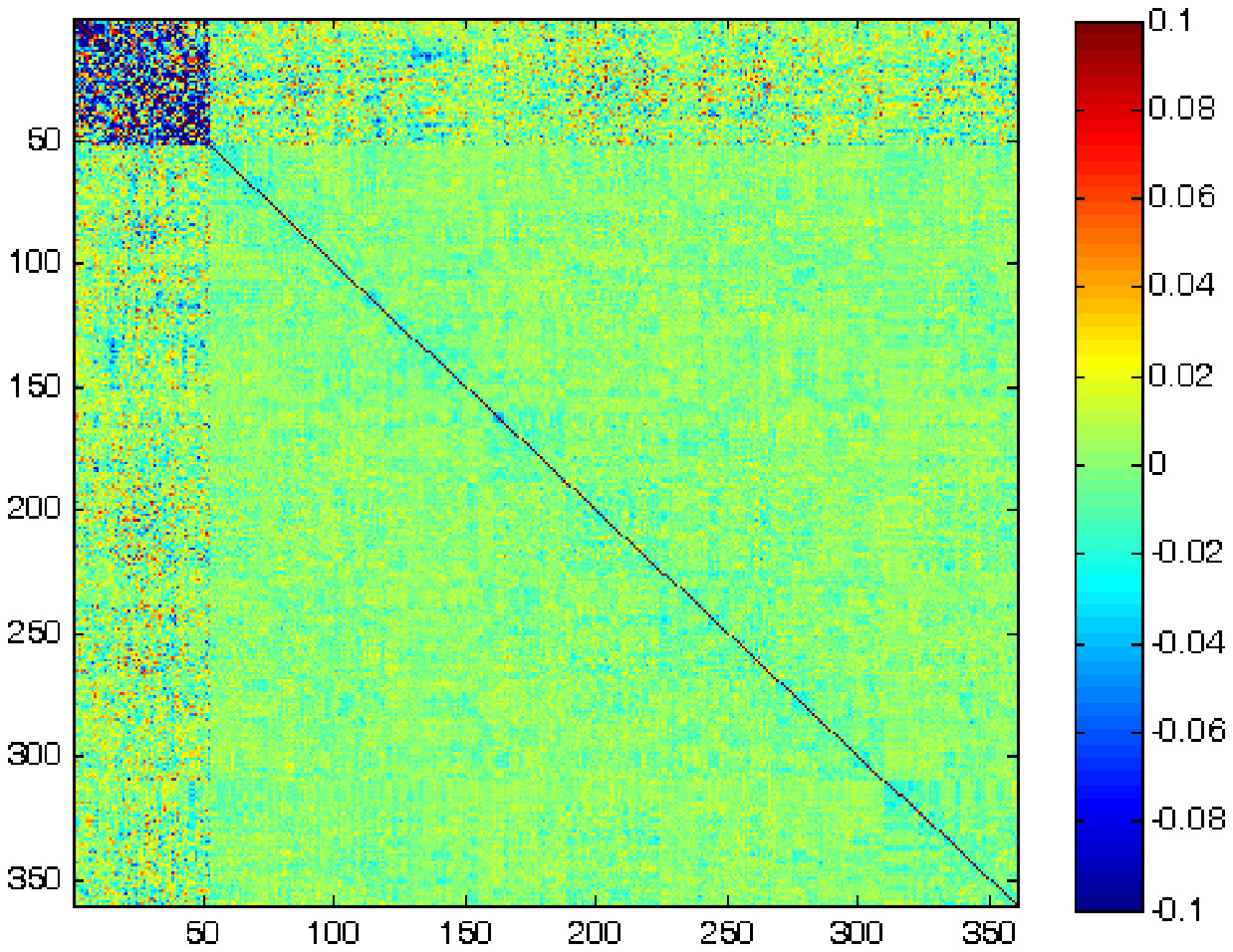}
\end{minipage}
\begin{minipage}{0.66\linewidth}
(c)\includegraphics[width=0.9\linewidth]{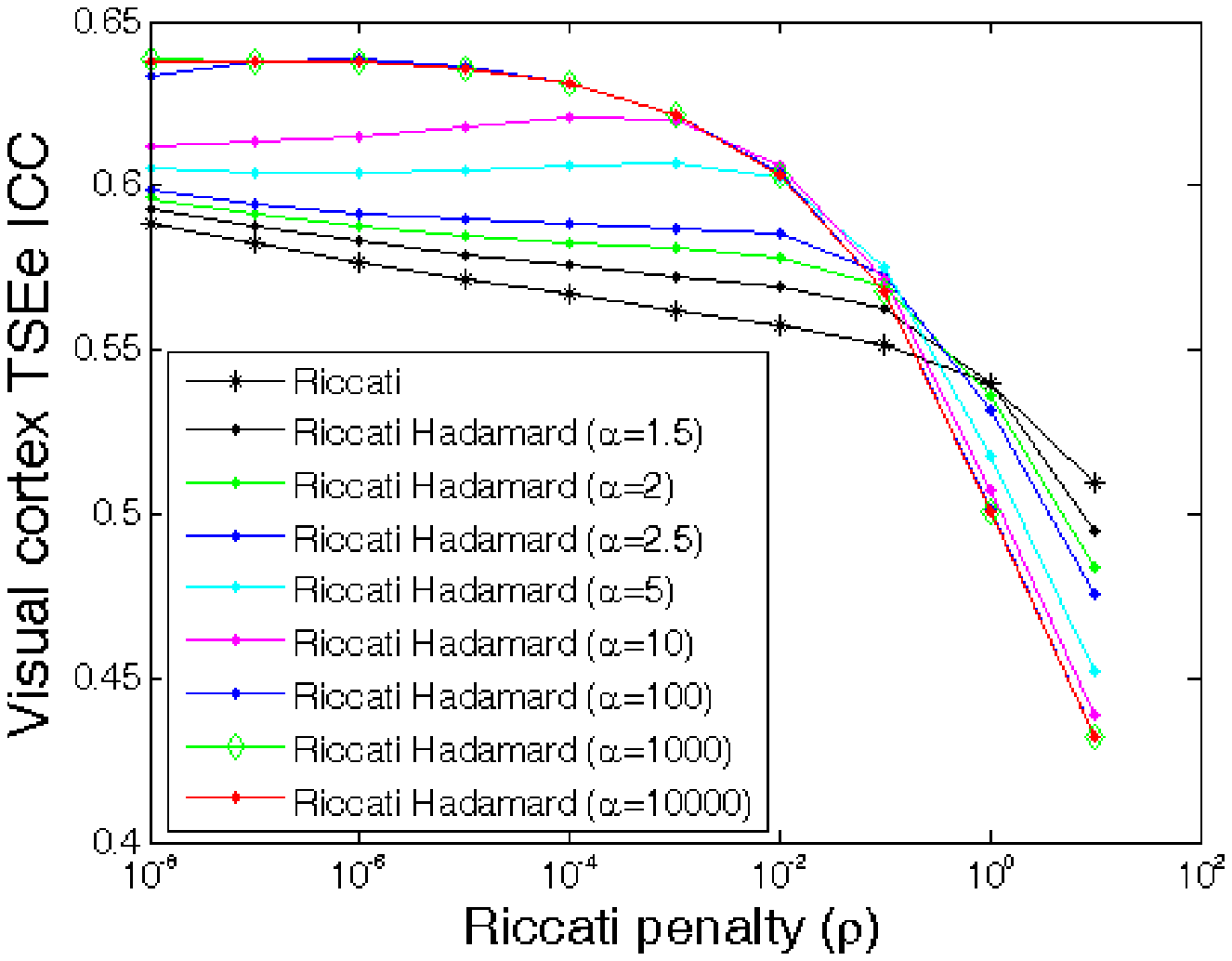}
\end{minipage}
\caption{\label{fig:exp:biomarkers} \textbf{Biomarkers extracted from Hadamard-Riccati precision matrices}. 
(a) Riccati regularized precision matrix 
(b) Hadamard Riccati regularized precision matrix
(c) Visual cortex TSEe ICC w.r.t Riccati penalties $\rho$ and non ROI suppression $\alpha$
}
\vspace{-1.0em}
\end{figure}

TSEe measures the integration of a functional subnetwork, and can, therefore, be considered as a biomarker. We observed that when TSEe is computed for Riccati regularized precision matrices, the test-retest reproducibility of this biomarker can sometimes be improved by penalizing the precisions involving nodes not part of the subnetwork of interest. During our experiments, the visual cortex was considered as the network of interest and we compared the ICC measured for different Riccati Hadamard penalizations. As explained in section \ref{method:biomarkers}, the vector $v$ defining the Hadamard Riccati penalty was set to $1$ for the nodes inside the visual cortex and $\alpha$ for the other nodes. The original Riccati penalty \cite{riccati_partial_correlations} corresponds to $\alpha=1$.
Figure \ref{fig:exp:biomarkers}.ab illustrates the effects of parameter $\alpha$. For large $\alpha$ values, the precisions outside the visual cortex are almost discarded and the Hadamard-Riccati penalization has the same effect as a restriction of the entire analysis to the visual cortex. This effect was beneficial in terms of biomarker ICC for small penalties, and detrimental for large penalties. 

\section{Discussion}


In this work, we have presented several neuroimaging applications of Riccati regularized precisions matrices. 
Because these precision matrices are low rank, stored efficiently, and the SVD required for their computation is fast, 
they can be computed at full brain resolution very efficiently, contrary to sparse precision matrices \cite{riccati_partial_correlations,varoquaux_nips}. 
However, we don't think that confronting these two approaches would be fully relevant. 
Sparse precision matrices elegantly capture the connectivity between brain regions, which is sparse by nature. 
By contrast, Riccati regularized matrices are designed for extracting the connectivity of large graphs where some homogeneity/redundancy 
is present and hence suitable for a low-rank description. We could claim that the first approach captures the integration of brain regions, whereas the second exploits the segregation of brain function. 
For this reason, we think that a combined framework, generating precisions matrices sparse for long range connections and low-rank for small range connections, should ideally leverage the benefits of both approaches. 

Because Riccati regularized and Tikhonov regularized 
precision matrices are computed in a similar fashion \cite{riccati_partial_correlations}, their main differences reside in the 
larger flexibility offered by the Riccati regularized matrices. 
Contrary to the Tikhonov penalization which acts only on the diagonal of the precision the Riccati regularization penalizes all the components 
of the matrix, which offers more freedom for designing penalties. 
A comparison of the eigenvalue transformation induced by the two penalties suggests also that the information corresponding to the large covariance eigenvalues is slightly better preserved into Riccati regularized precision matrices. 
The possibility of merging both penalties into a larger analytic framework is an interesting open question. 



The experiments presented in this paper have the potential to stimulate novel applications. 
For instance, similarly to section \ref{method:distance}, robust structural distances could be derived from the other cortical measures provided by Freesurfer \cite{freesurfer} such as areal distortion and cortical curvature, and for HCP myelin maps obtained by combining T1 and T2 weighted MRI scans \cite{hcp,hcp_preprocessing}. 
In addition, we emphasize that, by considering symmetric Riccati penalties only, we have restricted our investigations to optimization problems that can be solved efficiently but we have missed large families of applications. Asymmetric penalties would involve more elaborate algebraic Riccati equations and hopefully stimulate novel neuroimaging applications of control theory.

\section{Conclusion}

In this paper, we propose an integrated approach for the extraction of neuroimaging biomarkers. We measure the entropy of 
brain networks defined by computing Riccati penalized precision matrices. 
We demonstrate how these biomarkers can be improved by reducing data dimension via random projection. 
We highlight several neuroscience applications for which Riccati regularized precision matrices offer novel perspectives. 
These applications were all validated by processing the hundred unrelated subjects of the HCP dataset. 
We hope that the promising results obtained, both in terms of speed and test-retest performances, and the broad range of possible theoretical refinements 
will encourage further developments and additional neuroimaging applications. 


\subsection*{Acknowledgments}

Data were provided by the Human Connectome Project, WU-Minn Consortium (Principal Investigators: David Van Essen and Kamil Ugurbil; 1U54MH091657) funded by the 16 NIH Institutes and Centers that support the NIH Blueprint for Neuroscience Research; and by the McDonnell Center for Systems Neuroscience at Washington University

\bibliographystyle{splncs03}
\bibliography{biblio,biblio2} 

\end{document}